\documentclass[12pt]{article}
\begin{document}
\title{HOW FUNDAMENTAL IS GRAVITATION?}
\author{B.G. Sidharth\\
International Institute for Applicable Mathematics \& Information Sciences\\
Hyderabad (India) \& Udine (Italy)\\
B.M. Birla Science Centre, Adarsh Nagar, Hyderabad - 500 063 (India)}
\date{}
\maketitle
\begin{abstract}
Based on a Planck scale underpinning for the universe, we deduce an expression for the gravitational constant which exhibits it as a distributional effect over all the particles of the universe. This solves a long standing puzzle, the so called Weinberg formula which gives a microphysical parameter in terms of a cosmic parameter. This was also discussed on the basis of a cosmology that correctly predicted a dark energy driven accelerating universe - the linkage is now established.
\end{abstract}
\section{Introduction}
More than five thousand years ago, the Rig Veda repeatedly raised the question: ``How is it that \underline{though unbound} the sun does not fall down?''\\
This was a question that puzzled thinking man over the millennia. Indian scholars right up to Bhaskaracharya who lived about a thousand years ago believed in some attractive force which was responsible for keeping the celestial bodies from falling down.\\
The same problem was addressed by Greek thinkers about two thousand five hundred years ago. They devised transparent material spheres to which each of the celestial objects were attached - the material spheres prevented them from falling down.\\
Unfortunately it was this answer to the age old question, which held up further scientific progress till the time of Kepler, for even Copernicus accepted the transparent material spheres.\\
Kepler had a powerful tool in the form of the accurate observations of Tycho Brahe. He also had the advantage of the Indian numeral system, which via the Arabs reached Europe just a few centuries earlier. These lead him to his famous laws of elliptical orbits with definite periods corelated to distances from the Sun.\\
This couching of natural phenomena in the terse language of mathematical symbols that could be manipulated, was the beginning of modern science.\\
The important point was that the Greek answer to the problem of why heavenly objects do not fall down - the transparent material spheres were now demolished. The age old question of why celestial bodies do not fall down come back to haunt again. Kepler himself speculated about some type of a magnetic force between the Sun and the Planets, rather on the lines of earlier speculations in India.\\
It was Newton who provided the breakthrough.\\
To qote Hawking \cite{hawk},``{\it The Philosophiae Naturalis Principia Mathematica} by Isaac Newton, first published in Latin in 1687, is probably the most important single work ever published in the physical sciences. Its significance is equalled in the biological sciences only by {\it The Origin of Species}. The original impulse which caused Newton to write the {\it Principia} was a question from Edmund Halley as to whether the elliptical orbits of the planets could be accounted for on the hypothesis of an inverse square force directed towards the Sun. This was something that Newton had worked out some years earlier but had not published, like most of his work on mathematics and physics. However, Halley's challenge, and the desire to refute the suggestions of others such as Hooke and Descartes, spurred Newton to try to write a proper account of this result.''\\
Newton using Galileo's ideas of Mechanics, thus stumbled upon the Universal Law of Gravitation.\\
This held sway for nearly two hundred and twenty five years, before Einstein came out with his own theory of gravitation. There was no force in the mechanical sense that Newton and preceeding scholars had envisaged it to be. Rather it was due to the curvature of spacetime itself. Einstein's bizarre ideas have had some experimental verification while there are some other experimental consequences, such as gravitational waves, which need to be confirmed.\\
After Einstein's formulation of gravitation a problem that has challenged and defied solution has been that of providing a unified description of gravitation along with other fundamental interactions. One of the earliest attempts was that of Hermann Weyl, which though elegant was rejected on the grounds that in the final analysis, it was not really a unification of gravitation with electromagnetism but rather an adhoc prescription.\\
Modern approaches to this problem have finally lead to the abandonment of a smooth spacetime manifold. Instead, the Planck scale is now taken to be a minimum fundamental scale.\\
In earlier communications \cite{psu,ijmpa} we had argued from different points of view to arrive at the otherwise empirically known equations
$$R = \sqrt{N} l_P = \sqrt{N} l$$
\begin{equation}
l = \sqrt{n} l_P\label{e1}
\end{equation}
where $l_P, l$ and $R$ are the Planck length, the pion Compton wavelength and the radius of the universe and $N, \bar {N}$ and $n$ are certain large numbers. Some of these are well known empirically for example $\bar {N} \sim 10^{80}$ being the number of elementary particles, which typically are taken to be pions in the literature, in the universe.\\
One way of arriving at the above relations is by considering a series of $N$ Planck mass oscillators which are created out of the Quantum Vaccuum. In this case (Cf. also ref.\cite{ng}) we have
\begin{equation}
r = \sqrt{N a^2}\label{e2}
\end{equation}
In (\ref{e2}) $a$ is the distance between the oscillators and $r$ is the extent. Equations (\ref{e1}) follow from equation (\ref{e2}).\\
There is another way of arriving at equations (\ref{e1}) (Cf.ref.\cite{ijmpa}). For this, we observe that the position operator for the Klein-Gordan equation is given by \cite{sch},
$$\vec X_{op} = \vec x_{op} - \frac{\imath \hbar c^2}{2} \frac{\vec p}{E^2}$$
Whence we get
\begin{equation}
\hat X^2_{op} \equiv \frac{2m^3 c^4}{\hbar^2} X^2_{op} = \frac{2m^3 c^6}{\hbar^2} x^2 + \frac{p^2}{2m}\label{e3}
\end{equation}
It can be seen that purely mathematically (\ref{e3}) for $\hat {X}^2_{op}$ defines the Harmonic oscillator equation, this time with quantized, what may be called space levels. It turns out that these levels are all multiples of $(\frac{\hbar}{mc})^2$. This Compton length is the Planck length for a Planck mass particle. Accordingly we have for any system of extension $r$,
$$r^2 \sim Nl^2$$
which gives back equation (\ref{e1}). It is also known that the Planck length is also the Schwarzschild radius of a Planck mass, that is we have
\begin{equation}
l_P = Gm_P/c^2\label{e4}
\end{equation}
Using equations (\ref{e1}) and (\ref{e4}), we will now deduce a few new and valid and a number of otherwise empirically known relations involving the various microphysical parameters and large scale parameters. Some of these relations are deducible from the others. Many of these relations featured (empirically) in Dirac's Large Number Cosmology. We follow Dirac and Melnikov in considering $l, m, \hbar, l_P, m_P$ and $e$ as microphysical parameters \cite{dirac,melnikov}. Large scale parameters include the radius and the mass of the universe, the number of elementary particles in the universe and so on.\\
In the process we will also examine the nature of gravitation. It must also be observed that the Large Number relations below are to be considered in the Dirac sense, wherein for example the difference between the electron and pion (or proton) masses is irrelevant \cite{weinberg}.
\section{Interrelationships}
We will use the following well known equation which has been obtained through several routes:
\begin{equation}
\frac{GM}{c^2} = R\label{e5}
\end{equation}
For example in an uniformly expanding flat Friedman spacetime, we have \cite{weinberg}
$$\dot {R}^2 = \frac{8\pi G\rho R^2}{3}$$
If we substitute $\dot {R} = c$ at the radius of the universe in the above we recover (\ref{e5}).\\
We now observe that from the first two relations of (\ref{e1}), using the Compton wavelength expression we get
\begin{equation}
m = m_P/\sqrt{n}\label{e6}
\end{equation}
Using also the second relation in (\ref{e1}) we can easily deduce
\begin{equation}
N = \bar {N} n\label{e7}
\end{equation}
Using (\ref{e1}) and (\ref{e5}) we have
\begin{equation}
M = \sqrt{N} m_P\label{e8}
\end{equation}
Interestingly (\ref{e8}) can be obtained directly, without recourse to (\ref{e5}), from the energy of the Planck oscillators (Cf.ref.\cite{psu}). Combining (\ref{e8}) and (\ref{e6}) we get
\begin{equation}
M = \left(\sqrt{N n}\right) m\label{e9}
\end{equation}
Further if we use in the last of equation (\ref{e1}) the fact that $l_P$ is the Schwarzchild radius that is equation (\ref{e4}), we get, 
\begin{equation}
G = \frac{lc^2}{nm}\label{e10}
\end{equation}
We now observe that if we consider the gravitational energy of the $N$ Planck masses (which do not have any other interactions) we get,
$$\mbox{Gravitational \, Energy}\, = \frac{GNm^2_P}{R}$$
If this is equated to the inertial energy in the universe, $Mc^2$, as can be easily verified we get back (\ref{e5}). In other words the inertial energy content of the universe equals the gravitational energy of all the $N$ Planck oscillators.\\
Similarly if we equate the gravitational energy of the $n$ Planck oscillators constituting the pion we get
\begin{equation}
\frac{Gm^2_Pn}{R} = mc^2\label{e11}
\end{equation}
Using in (\ref{e11}) equation (\ref{e4}) we get
$$\frac{l_Pm_Pn}{R} = m$$
Whence it follows on using (\ref{e7}), (\ref{e6}) and (\ref{e1}),
\begin{equation}
n^{3/2} = \sqrt{N}, \, n = \sqrt{\bar {N}}\label{e12}
\end{equation}
Substituting the value for $n$ from (\ref{e12}) into (\ref{e10}) we will get
\begin{equation}
G = \frac{lc^2}{\sqrt{\bar {N}}m}\label{e13}
\end{equation}
If we use (\ref{e12}) in (\ref{e9}) we will get
\begin{equation}
M = \bar {N} m\label{e14}
\end{equation}
Alternatively we could use (\ref{e14}) which expresses the fact that the mass of the universe is given by the mass of the $\bar {N}$ elementary particles in it and deduce equations (\ref{e11}), (\ref{e12}) and (\ref{e13}). Using the expressions for the Planck length as a Compton wavelength and equating it to (\ref{e4}) we can easily deduce
\begin{equation}
Gm^2 = \frac{e^2}{n} = \frac{e^2}{\sqrt{\bar{N}}}\label{e15}
\end{equation}
wherein we have also used $\hbar c \sim e^2$ and (\ref{e6}). Equation (\ref{e15}) is another empirically well known equation which was used by Dirac in his Cosmology. Interestingly, as we have deduced (\ref{e15}), rather than use it empirically, this points to a unified description of electromagnetism and gravitation.\\
Interestingly also rewriting (\ref{e13}) as
$$G = \frac{l^2c^2}{Rm}$$
wherein we have used (\ref{e1}) and further using the fact that $H = c/R$, where $H$ is the Hubble constant we can deduce
\begin{equation}
m \approx \left(\frac{H\hbar^2}{Gc}\right)^{\frac{1}{3}}\label{e16}
\end{equation}
Equation (\ref{e16}) is the so called mysterious Weinberg formula, known empirically \cite{weinberg}. As Weinberg put it, ``...it should be noted that the particular combination of $\hbar , H, G$, and $c$ appearing (in the formula) is very much closer to a typical elementary particle mass than other random combinations of these quantities; for instance, from $\hbar , G$, and $c$ alone one can form a single quantity $(\hbar c/G)^{1/2}$ with the dimensions of a mass, but this has the value $1.22 \times 10^{22} MeV/c^2$, more than a typical particle mass by about $20$ orders of magnitude!\\
``In considering the possible interpretations (of the formula), one should be careful to distinguish it from other numerical ``coincidences''... In contrast, (the formula) relates a single cosmological parameter, $H$, to the fundamental constants $\hbar , G, c$ and $m$, and is so far unexplained.''\\
We will come back to this point but remark that (\ref{e13}) brings out gravitation in a different light-- somewhat on the lines of Sakharov. In fact it shows up gravitation as the excess or residual energy in the universe.\\
Finally it may be observed that (\ref{e13}) can also be rewritten as 
\begin{equation}
\bar {N} = \left(\frac{c^2 l}{mG}\right)^2 \sim 10^{80}\label{ex}
\end{equation}
and so also (\ref{e10}) can be rewritten as
$$n = \left(\frac{lc^2}{Gm}\right) \sim 10^{40}$$
It now immediately follows that
$$N \sim 10^{120}$$
Looking at it this way, given $G$ and the microphysical parameters we can deduce the numbers $N, \bar{N}$ and $n$.
\section{Comments}
Thus the many so called large number coincidences and the mysterious Weinberg formula can be deduced on the basis of a Planck scale underpinning for the elementary particles and the whole universe. This was done from a completely different point of view, namely using fuzzy spacetime and fluctuations in a 1997 model that successfully predicted a dark energy driven accelerating universe with a small cosmological constant \cite{ijmpa,cu}.\\
However the above treatment brings out the role of the Planck scale particles in the Quantum Vaccuum. It resembles, as remarked earlier the Sakharov-Zeldovich metric elasticity of space approach \cite{sakharov}. Essentially Sakharov argues that the renormalization process in Quantum Field Theory which removes the Zero Point energies is altered in General Relativity due to the curvature of spacetime, that is the renormalization or subtraction no longer gives zero but rather there is a residual energy similar to the modification in the molecular bonding energy due to deformation of the solids. We see this in a little more detail following Wheeler \cite{mwt}. The contribution to the Lagrangian of the Zero Point energies can be given in a power series as follows
$${\it L} (r) = A\hbar \int k^3 dk + B\hbar^{(4)} r \int k dk$$
$$+ \hbar [C(^{(4)} r)^2 + Dr^{\alpha \beta} r_{\alpha \beta}] \int k^{-1} dk$$
\begin{equation}
+ (\mbox{higher-order \, terms}).\label{e17}
\end{equation}
where $A, B, C$ etc. are of the order of unity and $r$ denotes the curvature. By renormalization the first term in (\ref{e17}) is eliminated. According to Sakharov, the second term is the action principle term, with the exception of some multiplicated factors. (The higher terms in (\ref{e17}) lead to corrections in Einstein's equations). Finally Sakharov gets
\begin{equation}
G = \frac{c^3}{16\pi B \hbar \int k dk}\label{e18}
\end{equation}
Sakharov then takes a Planck scale cut off for the divergent integral in the denominator of (\ref{e18}). This immediately yields
\begin{equation}
G \approx \frac{c^3 l^2_P}{\hbar}\label{e19}
\end{equation}
Infact using relations like (\ref{e1}), (\ref{e6}) and (\ref{e12}), it is easy to verify that (\ref{e19}) gives us back (\ref{e10}) (and (\ref{e13})).\\
According to Sakharov (and (\ref{e19})), the value of $G$ is governed by the Physics of Fields and Particles and is a measure of the metrical elasticity at small spacetime intervals. It is a microphysical constant.\\
However in our interpretation of (\ref{e13}) (which is apparently the same as Sakharov's equation (\ref{e19})), $G$ appears as the expression of a residual energy over the entire universe: The entire universe has an underpinning of the $N$ Planck oscillators and is made up of $\bar {N}$ elementary particles, which again each have an underpinning of $n$ Planck oscillators. It must be reiterated that (\ref{e19}) obtained from Sakharov's analysis shows up $G$ as a microphysical parameter because it is expressed in their terms. This is also the case in Dirac's cosmology.  This is also true of (\ref{e10}) because $n$ relates to the micro particles exclusively.\\
However when we use the relation (\ref{e12}), which gives $n$ in terms of $\bar {N}$, that is links up the microphysical domain to the large scale domain, then we get (\ref{e13}). With Sakharov's equation (\ref{e19}), the mysterious nature of the Weinberg formula remains. But once we use (\ref{e13}), we are effectively using the large scale character of $G$ -- it is not a microphysical parameter. This is brought out by (\ref{ex}), which is another form of (\ref{e13}). If $G$ were a microphysical parameter, then the number of elementary particles in the universe would depend solely on the microphysical parameters and would not be a large scale parameter. The important point is that $G$ relates to elementary particles and the whole universe \cite{bgs}. That is why (\ref{e13}) or equivalently the Weinberg formula (\ref{e16}) relate supposedly microphysical parameters to a cosmological parameter. Once the character of $G$ as brought out by (\ref{e13}) is recognized, the mystery disappears.
\section{Fluctuations I}
In 1997 a model \cite{ijmpa} correctly predicted a dark energy driven accelerating universe with a small cosmological constant at a time when the belief was that, aided by dark matter, the universe was slowing down. This contra view was confirmed by observational evidence in 1998 and thereafter. This model also deduced from theory some well known empirical relations between cosmological parameters and constants from micro physics, relations which had puzzled astronomers for nearly a century.\\
Cosmic parameters include the radius of the universe $R$ and the Hubble constant $H$  and microphysical parameters include $\hbar, c$ and $l$ and $m$ (the Compton wavelength and mass of a typical elementary particle, taken in the literature to be a pion) and $e$. They have been considered to be puzzling coincidences. For example we have,
\begin{equation}
R = \sqrt{N} l\label{ea1}
\end{equation}
\begin{equation}
\frac{Gm^2}{e^2} = \frac{1}{\sqrt{N}} \sim 10^{-40}\label{ea2}
\end{equation}
or the so called Weinberg formula
\begin{equation}
m = \left(\frac{H \hbar^2}{Gc}\right)^{\frac{1}{3}}\label{ea3}
\end{equation}
where $N \sim 10^{80}$ is the number of elementary particles, typically pions, in the universe. Equation (\ref{ea1}) is the well known Weyl-Eddington formula. On the other hand (\ref{ea2}) which is the ratio of the electromagnetic and Gravitational coupling constants, is deducible from (\ref{ea3}). The very mysterious feature of (\ref{ea3}) has been stressed by Weinberg as we saw earlier.\\
Relations like (\ref{ea1}) and (\ref{ea2}) inspired the Dirac Large Number Cosmology \cite{ijmpa,dirac}. All these relations are to be taken in the order of magnitude sense.\\
We will now take a different route and provide a theoretical rational for equations (\ref{ea1}), (\ref{ea2}) and (\ref{ea3}), and in the process light will be shed on the new cosmological model and the nature of gravitation as we saw earlier.
\section{Large Number Relations}
Following Sivaram \cite{sivaram} we consider the gravitational self energy of the pion. This is given by
$$\frac{Gm^2}{l} = Gm^2/(\hbar/mc)$$
If this energy were to have a life time of the order of the age of the universe, $T$, then we have by the Uncertainty relation
\begin{equation}
\left(\frac{Gm^3c}{\hbar}\right) (T) \approx \hbar\label{ea4}
\end{equation}
As $T = \frac{1}{H}$, (\ref{ea4}) immediately gives us the Weinberg formula (\ref{ea3}). It must be observed that (\ref{e4}) gives a time dependent gravitational constant $G$.\\
We could also derive (\ref{ea3}) by an argument given by Landsberg \cite{land}. We use the fact that the mass of a particle is given by
\begin{equation}
m(b) \sim \left(\frac{\hbar^3 H}{G^2}\right)^{1/5} \left(\frac{c^5}{\hbar H^2 G}\right)^{b/15}\label{ea5}
\end{equation}
where $b$ is an unidentified constant. Whence we have
$$m(b) \sim G^{-3/5} G^{-3b/15} = G^{-(b+1)/5}$$
The mass that would be time independent, if $G$ were time dependent would be given by the value
$$b = -1$$
With this value of $b$ (\ref{ea5}) gives back (\ref{ea3}).\\
However, let us proceed along a different track. We rewrite (\ref{e4}) as
\begin{equation}
G = \frac{\hbar^2}{m^3c} \cdot \frac{1}{T}\label{ea6}
\end{equation}
If we use the fact that $R = cT$, then (\ref{ea6}) can be written as
\begin{equation}
G = \frac{\hbar^2}{m^3R}\label{ea7}
\end{equation}
Let us now use the well  known relation \cite{nottale,ruffini,cu}
\begin{equation}
R = \frac{GM}{c^2},\label{ea8}
\end{equation}
There are several derivations of (\ref{ea8}). For example in a uniformly expanding Friedman universe (Cf.ref.\cite{weinberg}), we have
$$\dot {R}^2 = \frac{8\pi G \rho R^2}{3}$$
If we substitute the value $\dot {R} = c$ at the radius of the universe, then we recover (\ref{ea8}). If we use (\ref{ea8}) in (\ref{ea7}) we will get
\begin{equation}
G^2 = \frac{\hbar^2 c^2}{m^3 M}\label{ea9}
\end{equation}
Let $M/m = N$ be called the number of elementary particles in the universe. Then (\ref{ea9}) can be written as
\begin{equation}
G = \frac{\hbar c}{m^2 \sqrt{N}}\label{ea10}
\end{equation}
We observe that (\ref{ea10}) can also be written as
\begin{equation}
Gm^2/e^2 \sim \frac{1}{\sqrt{N}}\label{ea11}
\end{equation}
Using (\ref{ea10}) in (\ref{ea8}) it is easy to verify that we get
\begin{equation}
\sqrt{N} l = R\label{ea12}
\end{equation}
We now observe that (\ref{ea12}) is the same as (\ref{ea1}) while (\ref{ea11}) is the same as (\ref{e2}), with $N$ taken to be $\sim 10^{80}$, the well known number of elementary particles in the universe. Moreover (\ref{ea10}) becomes, on using (\ref{ea12}) and the fact that $H = \frac{c}{R}$, the mysterious Weinberg formula (\ref{ea3}).\\
We now remark that (\ref{ea6}) shows an inverse dependence on time of the gravitation constant, while (\ref{ea10}) shows an inverse dependence on $\sqrt{N}$. Equating the two, we get,
\begin{equation}
T = \sqrt{N} \tau\label{ea13}
\end{equation}
another well known Large Number relation. If we now take the time derivative of (\ref{ea10}) and use (\ref{ea13}), we get
\begin{equation}
\dot {N} = \frac{\sqrt{N}}{\tau}\label{ea14}
\end{equation}
Equation (\ref{ea14}) is the starting point of the fluctuational cosmology referred to (Cf.ref.\cite{ijmpa}). To put it briefly in a phase transition from the Quantum Vaccuum $\sqrt{N}$ particles appear within the Compton time $\tau$. Starting from (\ref{ea14}) it is possible to deduce the various relations (\ref{e1}) to (\ref{ea3}) and (\ref{ea6}), (\ref{ea10}) and (\ref{ea13}). It was shown that this leads to an ever expanding universe with a Hubble constant given by,
$$H = \frac{Gcm^3}{\hbar^2},$$
which is the same as (\ref{ea3}), and a small cosmological constant given by
$$\wedge \leq 0 (H^2)$$
One of the problems with other cosmological models has been an embarassingly large $\wedge$ which does not agree with observation.
\section{Remarks}
We now make a few remarks. Firstly it is interesting to note that $\sqrt{N}m$ will be the mass added to the universe. Let us now apply the well known Beckenstein formula for the life time of a mass $M$ viz., \cite{ruffini},
$$t \approx G^2M^3/\hbar c^4$$
to the above mass. The life time as can be easily verified turns out to be exactly the age of the universe!\\
The second remark is that (\ref{ea10}) shows that $G$ has a distributional character over all the $N$ particles in the universe, that is it is not a microphysical parameter \cite{bgs}. This now explains the mystery of the Weinberg formula (\ref{ea3}) - the presence of $G$ along with $H$ shows that there are two large scale parameters in (\ref{ea3}). In other words (\ref{ea3}) does not give a single cosmological parameter in terms of purely microphysical parameters.\\
We next observe that with the dependence on time of $G$, given by, for example (\ref{ea6}), it is possible to recover standard effects like the precession of the perihelion of Mercury or the bending of light \cite{cu,nc}.\\
A Final remark. To appreciate the role of fluctuations in the otherwise mysterious Large Number relations, let us follow Hayakawa \cite{hayakawa} and consider the excess of electric energy due to the fluctuation $\sim \sqrt{N}$ of the elementary particles in the universe and equate it to the inertial energy of an elementary particle. We get
$$\frac{\sqrt{N}e^2}{R} = mc^2$$
This gives us back (\ref{ea2}) if we use (\ref{ea8}). If we use (\ref{ea1}) on the other hand, we get
$$e^2/mc^2 = l,$$
another well known relation from micro physics.
\section{Fluctuations II}
We start with the current view of Planck scale oscillators in the background dark energy or Quantum Vaccuum. In this context it has been shown that elementary particles can be considered to be normal modes of $n \sim 10^{40}$ Planck oscillators in the ground state, while the etire universe itself has an underpinning of $N \sim 10^{120}$ such oscillators, there being $\bar {N} \sim 10^{80}$ elementary particles in the universe \cite{psu}. These Planck oscillators are formed out of the Quantum Vaccuum (or dark energy). Thus we have, $m_P c^2$ being the energy of each Planck oscillator, $m_P$ being the Planck mass, $\sim 10^{-5}gms$,
\begin{equation}
m = \frac{m_P}{\sqrt{n}}, l = \sqrt{n} l_P, \tau = \sqrt{n} \tau_P , n = \sqrt{\bar{N}}\label{eb1}
\end{equation}
where $m$ is the mass of a typical elementary particle, taken to be a pion in the literature. In the sequel we will denote the mass, Compton length and Compton time of an elementary particle like the pion by $m, l$ and $\tau$ while the same symbols with subscript $P$ denote the Planck mass and Planck scale.\\
Similarly the ground state of $N$ such Planck oscillators would be
\begin{equation}
\bar {m} = \frac{m_P}{\sqrt{N}} \sim 10^{-65}gms\label{eb2}
\end{equation}
The universe is an excited state and consists of $N$ such ground state levels and so we have, from (\ref{eb2})
$$M = \bar{m} N = \sqrt{N} m_P \sim 10^{55}gms,$$
as required, $M$ being the mass of the universe.\\
Due to the fluctuation $\sim \sqrt{n}$ in the levels of the $n$ oscillators making up an elementary particle, the energy is, remembering that $mc^2$ is the general state,
$$\Delta E \sim \sqrt{n} mc^2 = m_P c^2,$$
using (\ref{eb1}), and so the indeterminacy time is
$$\frac{\hbar}{m_Pc^2} = \tau_P,$$
as indeed we would expect.\\
The corresponding minimum indeterminacy length would therefore be $l_P$. One of the consequences of the minimum spacetime cut off is that the Heisenberg Uncertainty Principle takes an extra term \cite{mup}. Thus we have 
\begin{equation}
\Delta x \approx \frac{\hbar}{\Delta p} + \alpha \frac{\Delta p}{\hbar},\, \alpha = l^2 (\mbox{or}\, l^2_P)\label{eb6}
\end{equation}
where $l$ (or $l_P$) is the minimum interval under consideration.
The first term gives the usual Heisenberg Uncertainty Principle.\\
Application of the time analogue of (\ref{e6}) for the indeterminacy time $\Delta t$ for the fluctuation in energy $\Delta \bar{E} = \sqrt{\bar{N}} mc^2$ in the $\bar{N}$ particle states of the universe gives exactly as above,
$$\Delta t = \frac{\Delta E}{\hbar} \tau^2_P = \frac{\sqrt{N}mc^2}{\hbar} \tau^2_P = \frac{\sqrt{N} m_Pc^2}{\sqrt{n}\hbar} \tau^2_P = \sqrt{n} \tau_P = \tau ,$$
wherein we have used (\ref{eb1}). In other words, for the fluctuation $\sqrt{\bar{N}}$, the time is $\tau$. It must be emphasized that the Compton time $\tau$ of an elementary particle, is an interval within which there are unphysical effects like zitterbewegung - as pointed out by Dirac, it is only on averaging over this interval, that we return to meaningful Physics.  This gives us,
\begin{equation}
d\bar{N}/dt = \sqrt{\bar{N}}/\tau\label{eb3}
\end{equation}
On the other hand due to the fluctuation in the $\sqrt{N}$ oscillators constituting the universe, the fluctuational energy is similarly given by
$$\sqrt{N} \bar {m} c^2,$$
which is the same as (\ref{eb2}) above. Another way of deriving (\ref{eb3}) is to observe that as $\sqrt{n}$ particles appear fluctuationally in time $\tau_P$ which is, in the elementary particle time scales, $\sqrt{n} \sqrt{n} = \sqrt{\bar{N}}$ particles in $\sqrt{n} \tau_P = \tau$. That is, the rate of the fluctuational appearance of particles is
$$
\left(\frac{\sqrt{n}}{\tau_P}\right) = \frac{\sqrt{\bar{N}}}{\tau} = d \bar{N}/dt$$
which is (\ref{eb3}).\\
Equation (\ref{eb3}) was used in the model of fluctuational cosmology which correctly predicted in advance a dark energy driven accelerating universe with a small cosmological constant, and also deduced from theory the supposedly mysterious and inexplicable large number relations including the hitherto puzzling and inexplicable Weinberg formula to be seen below \cite{cu,weinberg,ijmpa}. Thus it is possible to understand the fluctuations in terms of the underpinning of Planck scale oscillators in the Quantum Vaccuum.\\
We would now like to make some remarks. Starting from a completely different point of view namely Black Hole Thermodynamics, Landsberg \cite{land} deduced that the smallest observable mass in the universe is $\sim 10^{-65}gms$, which is exactly the minimum mass given in (\ref{eb2}).\\
Further due to the fluctuational appearance of $\sqrt{\bar{N}}$ particles, the fluctuational mass associated with each of the $\bar{N}$ particles in the universe is
$$\frac{\sqrt{\bar {N}} m}{\bar {N}} =  \frac{m}{\sqrt{\bar {N}}} \sim 10^{-65}gms,$$
that is once again the smallest observable mass or ground state mass in the universe.\\
We next remark that it was argued using the geometrical model of Santamoto that cosmic fluctuations could yield the Bohmian Quantum Mechanics \cite{santa1,santa2,santa3,santa4}. Indeed another way of seeing this is as follows: The fluctuation in the gravitational energy of a typical elementary particle is given by, following Hayakawa \cite{hayakawa,csfqmcf},
$$(\Delta mc^2)  = \frac{G\sqrt{\bar {N}}m^2}{R} = \frac{G\sqrt{\bar {N}}m^2}{c} \cdot \frac{1}{T},$$
or, we have,
\begin{equation}
(\Delta mc^2) T = \frac{G\sqrt{\bar {N}} m^2}{c} (= \hbar )\label{eb4}
\end{equation}
Interestingly, as can be verified the right side of (\ref{e4}) is the reduced Planck constant, while (\ref{eb4}) itself is an expression of the Uncertainty Principle.\\
Equally we could argue that in the random motion of $\bar {N}$ particles, $l$ the fluctuation in the length is given by
\begin{equation}
l \approx \frac{R}{\sqrt{\bar {N}}}\label{eb5}
\end{equation}
Specializing to the case of the universe where $l$ is the Compton length and $R$ is the radius of the universe, (\ref{eb5}) is the well known Weyl-Eddington formula, one of the mysterious large number relations. Incidentally while Sivaram has argued that $l$ plays the role of a fundamental length in strong and electromagnetic interactions, the existence of such a fundamental length leads to, in modern approaches a non commutative geometry with interesting consequences \cite{cu,mup,sivaram}.\\
 We will now relate the second term in (\ref{eb6}) with $l$ in place of $l_P$, to large scale fluctuations $\sim \sqrt{\bar {N}}$, in the universe, where $\bar {N} \sim 10^{80}$ is the number of elementary particles encountered earlier. Infact the second term gives with $\Delta x = R$ the radius of the universe,
$$\Delta x = R = \frac{Gm\bar{N}}{c^2} = \frac{\Delta p}{\hbar} l^2 = \frac{\sqrt{\bar{N}}mc}{\hbar} l^2$$
wherein we have used the well known relation \cite{nottale},
$$R = \frac{GM}{c^2} \left(= \frac{Gm\bar{N}}{c^2}\right),$$
and where the total uncertainty in momentum is $\sqrt{\bar{N}}mc$. 
Whence we have
\begin{equation}
G = \frac{c^3l^2}{\hbar \sqrt{\bar{N}}}\label{eb7}
\end{equation}
Interestingly (\ref{eb7}) which is the same as (\ref{ea10}) is equivalent to (\ref{eb4}) the right side of which is $\hbar$ as remarked. Even more interestingly (\ref{eb7}) is identical to the mysterious Weinberg formula
\begin{equation}
m \approx \left(\frac{H\hbar^2}{Gc}\right)^{\frac{1}{2}}\label{eb8}
\end{equation}
Equation (\ref{eb8}) as we saw,  has been a big puzzle because it relates a cosmological parameter $H$ with what have been considered to be microphysical constants.\\
However it can be seen that there is an explanation in large scale fluctuations and the related fuzzyness in spacetime leading to (\ref{eb7}) which is another form of (\ref{eb8}) (Cf. also \cite{bgs}). The point is, that (\ref{eb7}) shows the distributional character of $G$ over all the $\bar{N}$ particles of the universe, rather than as a microphysical constant. Thus the Weinberg formula is no longer mysterious, and moreover it is not accidental, but has been derived.  Equally interesting is the fact that (\ref{eb4}) or (\ref{eb7}) or (\ref{eb8}) give back
\begin{equation}
e^2/Gm^2 \sim \sqrt{\bar{N}} \approx 10^{40}\label{eb9}
\end{equation} 
which as we saw is another well known supposedly accidental relationship expressing the relative strengths of the electromagnetic and gravitation strengths.\\
It must be mentioned that all these ``Large Number Relations'' have been deduced alternatively on the basis of the fluctuational cosmology referred to earlier, whilch uses equation (\ref{eb3}).\\
So we can see the convergence of considerations from this approach and the apparently different approach of fuzzy spacetime and the Modified Uncertainty Principle. 
\section{Characterizing the Weak Interaction}
In earlier work, we had given a characterization of the weak interaction \cite{cu}. We will now examine this issue afresh. As noted, the balance of the gravitational force and the fermi energy of the cold background nutrinos gives \cite{hayakawa}
$$\frac{GN_\nu m^2_\nu}{R} = \frac{N^{2/3}_\nu \hbar^2}{m_\nu R^2},$$
Whence, $N_\nu$ the number of nutrinos $\sim 10^{90}$, as is known to be the case. We then use the fact that due to the fluctuation in the number of nutrinos, we have
$$\frac{\bar g^2 \sqrt{N_\nu}}{R} \approx m_\nu c^2$$
where $\bar g^2$ gives the weak interaction coupling constant. Comparing this with a similar equation for the electron we get \cite{cu}
$$\bar g^2 /e^2 \sim 10^{-13}$$
as indeed is known to be the case. To proceed further if as in earlier work, we equate the gravitational energy of the neutrino vis-a-vis other neutrinos with its inertial energy we get
$$\frac{GN_\nu m^2_\nu}{R} = m_\nu c^2$$
Whence we have
$$m_\nu \sim 10^{-8} m_e$$
which infact was confirmed by the Superkamiokande experiments. Thus it is possible to characterize the weak interactions in terms of fluctuations.\\
Interestingly $l$ the pion Compton wavelength turns out to be a fundamental length, as in the Heisenberg unified theory. Infact we have for the strong interaction
$$g^2_S / l = m_q c^2,$$
$m_q$ being the quark mass while for the electromagnetic interaction
$$e^2/l = mc^2$$
For the gravitational interaction, the coupling constant is given by $g^2_G = Gm^2$, and the minimum gravitational mass is given by as we saw earlier $m_G \sim 10^{-65}gms$. Whence we have
$$g^2_G / l = m_G c^2$$
We finally observe that if at all there is a smallest scale in the universe, then this scale is not smaller than the Planck scale. This is because, when the smallest scale is $a$, we have the relation
$$[x,p] = \imath \hbar \left(1 + \left(\frac{a}{\hbar}\right)^2 p^2\right)$$
which leads to the Modified Uncertainty Principle
$$\Delta x \Delta p \geq \hbar + \frac{a^2}{\hbar} \Delta p^2$$
Using in the second term of the Modified Uncertainty Principle the fact that $\Delta p = Mc$ while $\Delta x \geq \frac{GM}{c^2}$, the Schwarzchild radius of the arbitrary mass $M$, we get immediately
$$\Delta x mc \geq \frac{a^2}{\hbar} m^2 c^2 = \frac{Gm^2}{c^2} c$$
$$\mbox{therefore}\quad \quad  a^2 \geq G \hbar / c^3 = l^2_P$$
which shows that the Planck length is the minimum possible length in the universe, as is to be otherwise expected.\\
This also means that a Black Hole cannot have a mass less than the Planck mass $10^{-5}gms$. This is surprising, because in the usual theory of Black Holes, any arbitrary mass can be a Black Hole with a suitably defined Schwarzchild radius.

\end{document}